\documentstyle[multicol,prb,aps]{revtex}

\input{epsf}

\begin{document}
\draft

\title{The frequency, temperature, and magnetic field dependence 
of ferromagnetic resonance and anti-resonance in
La$_{0.8}$Sr$_{0.2}$MnO$_3$}

\author{Andrew Schwartz,\cite{email} Marc Scheffler,\cite{ms-address}
and Steven M. Anlage}
\address{MRSEC and Center for Superconductivity Research, 
Department of Physics, University of Maryland \\College Park, MD 20742-4111}
\date{lsmo4.tex, \today}
\maketitle

\begin{abstract}
Employing a broadband microwave reflection configuration, we have 
measured the complex surface impedance, $Z_S(\omega,T,H)$, of single 
crystal La$_{0.8}$Sr$_{0.2}$MnO$_3$, as a function of frequency 
(0.045--45~GHz), temperature (250--325~K), and magnetic field (0--1.9~kOe). 
The microwave surface impedance depends not only on the resistivity of 
the material, but also on the magnetic permeability, $\hat\mu(\omega,T,H)$,
which gives rise to ferromagnetic resonance (FMR) and ferromagnetic 
anti-resonance (FMAR). The broadband nature of this experiment allows
us to follow the FMR to low frequency and to deduce the behavior of
both the local internal fields and the local magnetization in the 
sample.
\end{abstract}

\pacs{PACS numbers: 76.50.+g,78.70.Gq,75.30.Vn,76.60.Jx}

\begin{multicols}{2}
\columnseprule 0pt
\narrowtext

\section{Introduction}
\label{sec:intro}

The recent discovery of large negative magnetoresistance in the 
manganite perovskites La$_{1-x}$A$_x$MnO$_3$ (where A is a divalent 
cation, typically Ca, Sr, or Ba) has led 
to numerous studies of both the physics and potential technological 
applications of these materials.\cite{vonHelmolt93,Jin94,Ramirez97}
A great deal of work has been done on the dc transport \cite{vonHelmolt93,Jin94,Ramirez97,Urushibara95,Mahendiran96}
and optical\cite{Kim98,Quijada98,Boris98} properties of these so-called
colossal magnetoresistive (CMR) oxides. But by 
comparison the microwave properties have not been extensively 
investigated, and the work that has been done has generally been limited 
to a few frequencies by the resonant techniques employed.
\cite{Lofland96,Lofland97a,Lofland97b,Lofland97c,Ramos98} In addition, 
these measurements generally do not produce quantitative values of the 
surface resistance or the surface reactance. 

Because of the relatively high resistivities of the CMR compounds, the
enhanced sensitivity of resonant microwave techniques is not necessary.
In this paper we present the 
results of our broadband, non-resonant microwave surface impedance 
measurements, in which we have quantitatively determined the complex 
surface impedance, $Z_S(\omega,T,H)=R_S(\omega,T,H)+iX_S(\omega,T,H)$, of La$_{0.8}$Sr$_{0.2}$MnO$_3$ single 
crystals over three decades in frequency, and as a function of both 
temperature and applied external magnetic field. We previously reported 
measurements of the surface impedance of this material in zero applied field,
and showed that it is possible to extract the temperature dependence of the
spontaneous magnetization from these data.\cite{Schwartz00} Here we focus on
the surface impedance in an applied magnetic field, and show that the spectra 
can be well described by the Landau-Lifshitz-Gilbert expression for the 
dynamic susceptibility of a ferromagnetic system. We present quantitative
analysis of the field and temperature dependence of the ferromagnetic resonance
and anti-resonance features observed in the spectra, and demonstrate that
it is possible to extract the local magnetization and internal fields
from these data.

La$_{0.8}$Sr$_{0.2}$MnO$_3$ (LSMO) has a ferromagnetic phase 
transition with a Curie temperature $T_C$ of approximately
305.5~K.\cite{Schwartz00} It is well established that the low temperature 
phase is a ferromagnetic metal (the resistivity drops dramatically below $T_C$), 
while above $T_C$ the system is paramagnetic, and the resistivity exhibits 
a negative slope with respect to temperature. The resistivity has a maximum
around $T_p=318$~K, significantly above $T_C$, which is typical in these 
manganite materials.\cite{Lofland97b} 

\section{Experimental setup}
\label{sec:expt}

In order to measure the temperature and frequency dependence of the 
complex surface impedance we have terminated a coaxial transmission line 
with the sample and measured its complex reflection coefficient. We have 
reported the details of this experimental geometry elsewhere,\cite{Booth94}
and will therefore give only a brief overview of 
the technique here. In particular, our original implementation of the 
method was for the study of superconducting thin films, so we will 
discuss its use here on bulk single crystals. 

This experiment is built around an HP8510C Vector Network Analyzer (NWA), 
which is continuously tunable over three decades from 45~MHz to 50~GHz. The 
phase-locked signal from the NWA is sent along a coaxial transmission line 
with a modified 1.85mm coaxial connector on the other end. This modified 
connector has a flat face and a spring-loaded center conductor, allowing 
a sample to be pressed against it, thereby terminating the transmission 
line. The amplitude and phase of the reflected signal are measured as the 
source is swept throughout the entire frequency range, and the complex 
reflection coefficient, $\hat{S}_{11}(\omega)$, is determined as the 
ratio of the reflected to incident voltages. After standard NWA 
calibration procedures to remove the errors due to the 
systematic response of the NWA and the 
transmission line, the complex surface impedance of the terminating 
material can be calculated from the reflection coefficient as follows: 
\begin{equation} 
\hat{Z}_S(\omega)=R_S(\omega)+iX_S(\omega)= 
Z_0\frac{1+\hat{S}_{11}(\omega)}{1-\hat{S}_{11}(\omega)}, 
\end{equation} 
where $Z_0=377\Omega$ is the impedance of free space. As discussed, due 
to the phase-sensitive detection capabilities of the NWA, it is possible 
to extract {\em both} the surface resistance, $R_S$, and the surface 
reactance, $X_S$, and the well-defined terminated transmission line 
geometry allows for {\em quantitative} evaluation of these material 
parameters. In addition, while this non-resonant technique does not 
provide the sensitivity of microwave resonator measurements, the 
broadband nature allows for the study of the response across three orders 
of magnitude in frequency, and therefore opens a unique window to 
dynamical processes within the material. 

By passing the coaxial line through a vacuum seal into a continuous flow 
cryostat, in which the sample is attached to the cold finger, we have 
measured the reflection coefficient as the temperature of the sample is 
varied. Because the temperatures of interest in this study were all above 
200~K, we operated the cryostat with a flow of liquid nitrogen rather 
than the traditional helium, and were able to achieve temperature 
stability on the order of a few millikelvin at all temperatures between 
200 and 325~K. This allowed us to study the temperature dependence of the 
surface impedance, to complement the frequency dependent data. 

As mentioned above, a standard calibration procedure is employed to 
remove systematic errors. In general there are three sources of error for 
which we must correct. One of these is due to the attenuation and phase 
delay of the coaxial transmission line between the NWA and the sample, 
and is therefore temperature dependent. Part of this line is within the 
cryogenic environment, and thus expands or contracts as the temperature 
is varied. Such changes in length cause changes in the phase delay and, 
to a lesser extent, the attenuation of the line. To correct for this, a 
further calibration procedure is performed. The temperature dependence of 
the reflection coefficient of 
a flat piece of bulk copper is measured as an additional calibration 
standard. It is measured in the same way (at the same temperatures and 
frequencies) as the sample. We then assume that this copper block acts as 
if it were a perfect short at all temperatures\cite{CuNote} and therefore 
has a known reflection coefficient ${\hat{S}_{11}^{\rm Cu}}=-1+0i$. 
Thus, corrected error coefficients are 
determined for all frequencies and temperatures of interest and the 
reflection coefficient of the sample can be deduced from the measured 
reflection coefficient.

In addition, by using two strong permanent magnets, one placed on either 
side of the sample, and varying the separation between them we have been 
able to produce a uniform static magnetic field at the sample position
which can be continuously varied 
from 0 to 1.9~kOe. The faces of the magnets are much larger than the size 
of the sample and therefore the field is expected to be relatively 
homogeneous throughout the sample. With this setup, we were able to study 
the field dependence of the surface impedance. For all of the data shown
in this paper, the field was taken to its maximum value of 1.9~kOe,
sufficient to saturate the magnetization, and then
the data were collected as the field was lowered to zero. 

\section{Ferromagnetic resonance and anti-resonance}
\label{sec:FMR}

In the presence of an applied external magnetic field of the appropriate 
magnitude, which will be discussed below, the microwave properties of 
ferromagnetic materials are characterized by two distinct features which 
result from the dispersion of the complex magnetic permeability 
$\mu(\omega)$:\cite{Lax62} ferromagnetic resonance (FMR) and 
ferromagnetic anti-resonance (FMAR). At the FMR frequency, $\omega_r$, 
the surface resistance, $R_S(\omega_r)$ shows a maximum due to a maximum 
in the imaginary part of the magnetic permeability, $\mu_2(\omega)$. At 
the same frequency, the real part of the permeability, $\mu_1(\omega)$, 
has a zero-crossing with negative slope. In order to satisfy the 
condition $\mu_1(\omega\rightarrow\infty)=1$, it is necessary that there 
be another zero-crossing, with positive slope, at a frequency 
$\omega_{ar}>\omega_r$. At this point, as will be shown below, the 
surface resistance shows a local minimum, commonly known as the 
ferromagnetic {\em anti}-resonance. This reduction in the
surface resistance is due to an increase in the penetration of the fields
into the material. It can also be shown that, in 
general, both $\omega_r$ and $\omega_{ar}$ depend not only on the 
externally applied field but also on the local internal magnetization of 
the material. Therefore, our measurements of the microwave surface 
impedance have yielded information about the magnetization of LSMO.

\subsection{Dynamic Susceptibility}
\label{sec:Chi}

In order to understand the origins of the ferromagnetic resonance and 
anti-resonance in more detail, and to be able to analyze the broadband
measurements presented here, it is necessary to examine the full functional 
form of the dynamic susceptibility $\chi(\omega)$.\cite{Lax62} The 
starting point for the calculation of the dynamic susceptibility of a 
ferromagnetic material is the Landau-Lifshitz equation of motion for the 
magnetization vector {\bf M} in the presence of a magnetic field {\bf H}:
\cite{Lax62}
\begin{equation}
\frac{d{\bf M}}{dt} = \gamma ({\bf M} \times {\bf H}) - 
\frac{\alpha\gamma}{M} ({\bf M} \times ({\bf M} \times {\bf H})),
\end{equation}
where $\gamma$ is the gyromagnetic ratio for an electron and $\alpha$ is 
a dimensionless damping parameter. The first term describes the 
precession of the magnetization around the applied field, and the second 
term describes the damping of this precessional motion. In this form, the 
damping term is perpendicular to {\bf M} and therefore changes only the 
angle and not the amplitude of the magnetization vector. This expression 
can be simplified further by taking the cross product of {\bf M} with 
both sides. After some algebraic manipulation this gives
\begin{equation}
\frac{d{\bf M}}{dt} = \gamma^* ({\bf M} \times {\bf H}) - 
\frac{\alpha}{M}\left[ {\bf M} \times \frac{d{\bf M}}{dt} \right],
\label{eq:final_LL}
\end{equation}
where $\gamma^*=\gamma(1+\alpha^2)$. In the limit of small damping 
($\alpha^2 \ll 1$), $\gamma^*\approx\gamma$. This is known as
the Landau-Lifshitz-Gilbert equation of motion, and $\alpha$ is
commonly referred to as the Gilbert damping parameter.

In order to calculate the dynamic susceptibility from this equation of 
motion, it is necessary to choose a form for the magnetic field and thus 
the magnetization. In the experiment we apply a uniform static magnetic 
field $H_0$, but the field which is important is the total {\em interal}
magnetic field $H_i$, which will in general differ from $H_0$, as we
will discuss below. So for now we just presume that there is a total static
internal field $H_i$ oriented along the $\hat{z}$ direction, and a 
microwave field ${\bf h}_{\rm rf,i}$, which may also differ from the
applied microwave field:
\begin{equation}
{\bf H}(t) = \hat{z}H_i + {\bf h}_{\rm rf,i} e^{i\omega t}. \label{eq:Htot}
\end{equation}
In an isotropic system, the magnetization is presumed to follow the 
applied field, so we assume that the static magnetization is aligned with 
the static field $H_i$ and that there is a component of the magnetization with 
the same time dependence as the microwave field:
\begin{equation}
{\bf M}(t) = \hat{z}M_0 + {\bf m}_{\rm rf} e^{i\omega t}. \label {eq:Mtot}
\end{equation}
By inserting Eqs.~(\ref{eq:Htot}) and (\ref{eq:Mtot}) into 
Eq.~(\ref{eq:final_LL}) we can easily calculate the susceptibility tensor 
$\overline{\overline{\chi}}$ from the relation
\begin{equation}
{\bf m}_{\rm rf}=\overline{\overline{\chi}} \cdot {\bf h}_{\rm rf,i} = \left( 
\begin{array}{ccc} 
\hat\chi_{xx} & \hat\chi_{xy} & 0\\
\hat\chi_{yx} & \hat\chi_{yy} & 0\\
0 & 0 & 0 
\end{array} 
\right)\cdot {\bf h}_{\rm rf,i},
\end{equation}
where we have assumed, as above, that the static field is in the 
$\hat{z}$ direction, and that ${\bf h}_{\rm rf,i}\ll{\bf H_i}$ and 
${\bf m}_{\rm rf}\ll{\bf M}$. It is clear here that there will be no components of 
${\bf m}_{\rm rf}$ along $\hat{z}$, and that any $\hat{z}$ component of 
${\bf h}_{\rm rf,i}$ will not contribute to ${\bf m}_{\rm rf}$. This is 
important for the geometry we have employed, and will be discussed 
further below.

Before calculating the components of the susceptibility tensor, however, 
it is necessary to account for the fact that in general 
the internal fields will differ from the applied external fields due to 
demagnetization effects. The applied external fields will lead to a net
magnetization, which in turn will
produce dipoles on the surface of the sample, and these dipoles give 
rise to a field within the sample which opposes the applied 
field.\cite{Lax62} The magnitude of such a demagnetization field ($H_d$) 
depends on the geometry of the sample and on the net magnetization. 
In addition, we will see later that 
it is necessary to allow for the existence of small static local
fields (${\bf H}_{loc}$) within the material. Such fields might arise, 
for example,
from anisotropy and domain structure. It is reasonable to assume that
$H_d$ will be aligned antiparallel to the applied 
field, but the orientation of ${\bf H}_{loc}$ is likely to vary with position 
within the sample. However, we will also presume that it is only the
component of ${\bf H}_{loc}$ aligned parallel to $M_0$ (i.e. along $\hat{z}$)
which will contribute to the precession of the magnetization. 
Thus we can write the total field within the material as follows:
\begin{equation}
{\bf H}_i =\hat{z}(H_0 - N_z M_0+H_{loc}).
\end{equation}
Similarly, the microwave magnetization produces a high frequency 
demagnetization field, so 
\begin{equation}
{\bf h}_{{\rm rf},i}={\bf h}_{\rm rf} - \hat{x}N_x^{\prime} m_x -
\hat{y}N_y^{\prime} m_y,
\label{eq:hrf}
\end{equation}
where ${\bf h}_{\rm rf}$ is the externally applied field, and we have 
omitted the term $\hat{z}N_z^{\prime} m_z$ because there will be no components of
${\bf m}_{\rm rf}$ along $\hat{z}$. $N_i$ are the geometry-dependent
demagnetization factors along the three principle axes and we have chosen
to denote dynamic factors in Eq.~(\ref{eq:hrf}) with primes because they
can, in general, differ from the static factors. We will
return to the precise determination of these values shortly.

Using these expressions we can write an expression for $\hat\chi_{xx}$, 
one of the nonzero diagonal components of the susceptibility tensor, as 
follows:
\begin{equation}
\hat\chi_{xx}(\omega) = \frac{\omega_M[(\omega_0+i\Gamma)+N_y^{\prime} 
\omega_M]} {\omega_r^2-(1+\alpha^2)\omega^2+i\Gamma[2\omega_0 + (N_x^{\prime}+N_y^{\prime})\omega_M ] }
\label{eq:Chi_components}
\end{equation}
where we have introduced the linewidth $\Gamma=\alpha\omega$. Similar 
expressions can be obtained for $\hat\chi_{yy}$, and for the off-diagonal 
terms $\hat\chi_{xy}$ and $\hat\chi_{yx}$,\cite{Scheffler98} but these 
will not be of importance here. In order to emphasize the dimensionless 
nature of the susceptibility and the fact that we have measured its 
frequency dependence, all of the fields have been expressed as frequencies, 
with the following definitions:
\begin{eqnarray}
\omega_M &=& \gamma^* \mu_0 M_0 \nonumber\\
\omega_0 &=& \gamma^* \mu_0 (H_0 - N_z M_0+H_{loc}) \nonumber \\
\omega_r &=& \sqrt{(\omega_0+N_x^{\prime}\omega_M)(\omega_0 + N_y^{\prime}\omega_M)}. \nonumber
\end{eqnarray}
It is clear from Eq.~(\ref{eq:Chi_components}) that for small damping 
($\alpha^2 \ll 1$), the quantity $\omega_r$ is the ferromagnetic 
resonance frequency. As written, these expressions are in SI units, with 
$H_0$ and $M_0$ in amperes-per-meter, all frequencies 
expressed in gigahertz, and the gyromagnetic ratio for the electron given 
by $\gamma/2\pi=28$~GHz/T.

In order to determine the form of the susceptibility it is only necessary 
to know the values of the demagnetization factors $N_i$ for the geometry 
of interest. For an arbitrarily-shaped sample, this is a difficult task, 
but for certain simple geometrical shapes, the demagnetization factors 
are well known. The samples of La$_{0.8}$Sr$_{0.2}$MnO$_3$ which we have 
investigated are disks, with diameters $d$ and thicknesses $t$ such that 
generally $t \leq 0.2 d$. The static field $H_0$ was applied in the plane
of the disk, thereby defining the $z$-axis. We take the $x$-axis to also
be in the plane, and the $y$-axis to be normal to the surface.
We can simplify the problem somewhat by taking into account the fact 
that the dynamic demagnetization factors ($N_x^{\prime}$ and $N_y^{\prime}$)
may be independent of the static one ($N_z$). In other words, the 
microwave fields penetrate only a short distance, the skin depth $\delta$, 
into the sample. For all measured frequencies, this penetration is much less than
the diameter of the sample, and therefore from the point of view of 
$h_{\rm rf}$, and therefore $m_{\rm rf}$, the sample looks like a 
thin plate, allowing us to set $N_x^{\prime}=0$ and $N_y^{\prime}=1$.
On the other hand, the static magnetic field 
penetrates throughout the sample, and therefore we cannot use this 
argument to determine $N_z$. However, in Sec.~\ref{sec:ResultsAnalysis}
we will show that we are able to extract a value of $N_z$ directly from
the microwave measurements.

As discussed above, the dynamic permeability, 
$\hat\mu=\mu_0(1+\hat\chi)$, has two distinct features: a resonance (FMR) 
at which $\mu_2$ is a maximum, and a zero-crossing of $\mu_1$ at higher 
frequency (FMAR). The frequencies of these two features are determined by 
the magnitudes of the total internal field $H_i$ and the local magnetization $M_0$, 
and by the demagnetization factors. With the applied field parallel to the plane 
of the sample, as in the measurements presented here, the FMR and 
FMAR frequencies are given by 
\begin{eqnarray} 
\omega_r &=& 
\gamma^*\mu_0\sqrt{H_i(H_i + M_0)} \label{eq:wr}\\ 
\omega_{ar} &=& \gamma^*\mu_0(H_i + M_0) \label{eq:war} 
\end{eqnarray}
where once again
we have taken $N_x^{\prime}=0$ and $N_y^{\prime}=1$. It is clear from 
Eqs.~(\ref{eq:wr}) and (\ref{eq:war}) that if we 
measure $\omega_r$ and $\omega_{ar}$ then we 
can uniquely determine both $M_0$ and $H_i$. 
It is worth noting that unlike in a typical magnetization measurement,
this magnetization $M_0$ is not the net magnetization of the sample
but instead is the average magnitude of the local magnetization.
Thus we see that by measuring the dynamic susceptibility it is possible 
to extract information about the {\em local} magnetization of the material. 
Such analysis will be discussed further in Sec.~\ref{sec:ResultsAnalysis}.

\subsection{Surface Impedance}
\label{sec:Zs}

In fact, as outlined above, the microwave reflection measurement which we 
have employed directly yields the surface impedance instead of the 
susceptibility, however the two are related through the permeability, 
$\hat\mu(\omega)=\mu_1(\omega)-i\mu_2(\omega)=\mu_0[1+\hat\chi(\omega)]$,
as follows:
\begin{equation}
\hat{Z}_S(\omega)=R_S(\omega)+iX_S(\omega)=\sqrt{i\omega\hat{\mu}(\omega)\rho_{dc}},
\label{eq:Zs}
\end{equation}
where $\rho_{dc}$ is the  $dc$ resistivity of the material. We have assumed that
at microwave frequencies La$_{0.8}$Sr$_{0.2}$MnO$_3$ is in the Hagen-Rubens 
limit (i.e. $\rho_2(\omega)\ll\rho_1(\omega) \approx \rho_{dc}$),
allowing us to insert a real 
frequency-independent value for $\rho$. Then we can substitute the 
expression for the susceptibility from Eq.~(\ref{eq:Chi_components}) into 
Eq.~(\ref{eq:Zs}) in order to model the complete frequency dependence of 
$R_S$ and $X_S$ and, in particular, the behavior in the vicinity of 
$\omega_r$ and $\omega_{ar}$. At the ferromagnetic resonance, $R_S$ is 
maximum and $X_S$ has an inflection point with negative slope, whereas at 
the anti-resonance, $R_S$ has a local minimum, while $X_S$ has another 
inflection point with positive slope.

In Sec.~\ref{sec:ResultsAnalysis} we will show that with a few 
small modifications to account for specifics of the measurement, 
Eq.~(\ref{eq:Zs}) gives an excellent description of the measured data. 
From such fits it is therefore possible to extract values of the local 
magnetization $M_0$ as a function of both temperature and applied 
magnetic field.

\section{Experimental Results and Analysis}
\label{sec:ResultsAnalysis}

Employing the configuration presented in Sec.~\ref{sec:expt} we have 
measured both the surface resistance and surface reactance of 
La$_{0.8}$Sr$_{0.2}$MnO$_3$ as functions of frequency, temperature, and 
applied magnetic field. We have measured the frequency dependence from 
45~MHz to 45~GHz but, as we will see, for the temperatures, 
magnetizations, and fields of interest in this paper the resonant 
features occur below 20~GHz, and we will therefore limit the discussion 
to that part of the frequency range. We have measured the temperature 
dependence of the response in the range 250--325~K, through the ferromagnetic 
transition at 305.5~K, and for fields of 0--1.9~kOe.

The single crystals of La$_{0.8}$Sr$_{0.2}$MnO$_3$ used in this study
were grown by the floating-zone technique\cite{Balbashov96} and 
the stoichiometry and structural 
integrity have been checked by x-ray diffraction and energy dispersive x-ray
analysis. From a 4~mm diameter rod, we cut disk-shaped samples with thicknesses
from 0.5--1~mm, and polished the surfaces to 1~$\mu$m flatness. 
Resistivity, ac susceptibility, and dc magnetization measurements have
been reported earlier on samples cut from the same boule.\cite{Lofland97b}

\subsection{Surface Impedance}

\begin{figure}[htb]
\begin{center}
\leavevmode
\epsfysize=13cm
\epsfbox{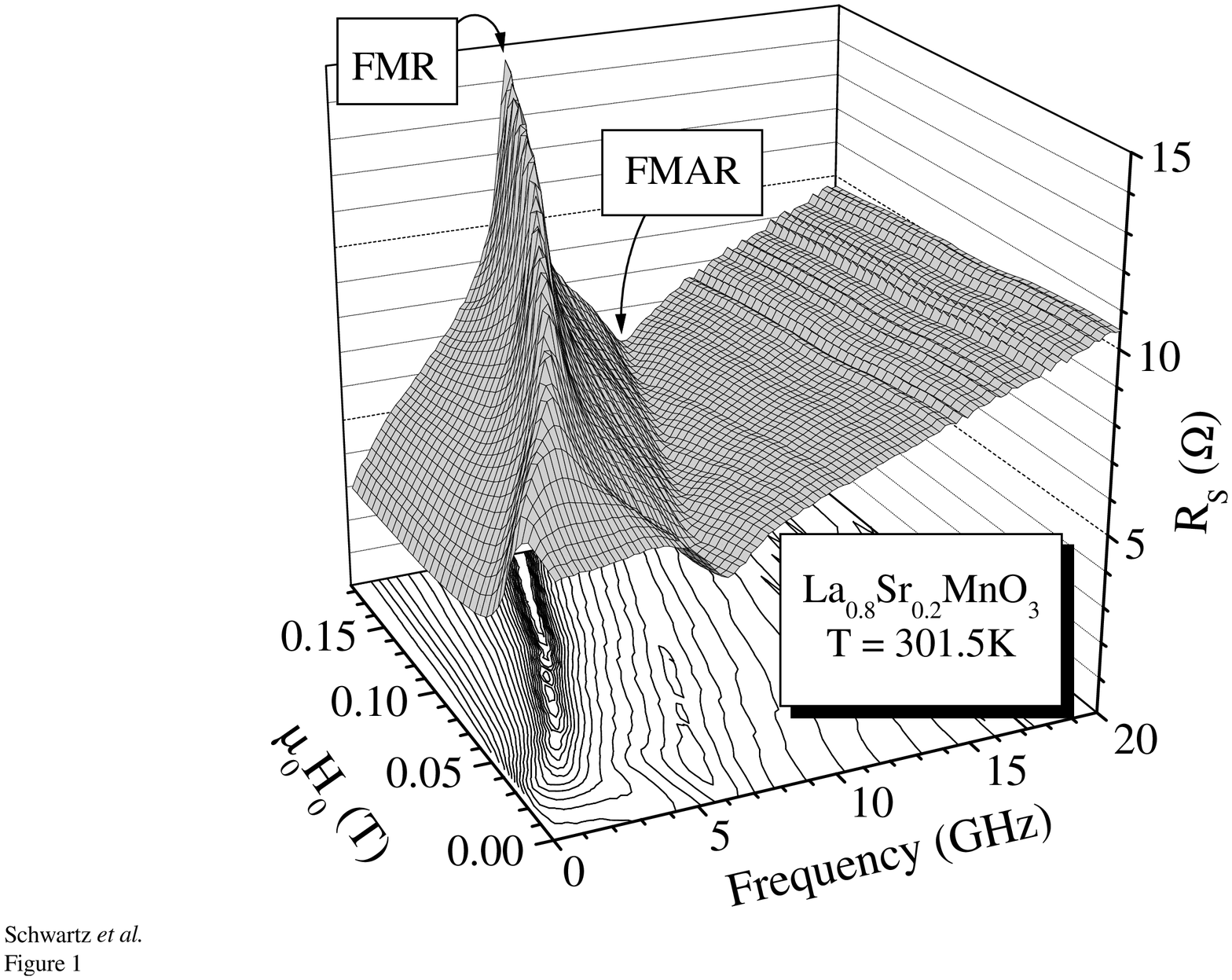}
\end{center}
\caption{The frequency and field dependence of the surface resistance 
$R_S$ of La$_{0.8}$Sr$_{0.2}$MnO$_3$ at $T=301.5$~K. As indicated, the 
ferromagnetic resonance (FMR) and ferromagnetic anti-resonance (FMAR) are 
clearly visible. The frequency and field dependencies of these two 
features are seen in the contour plot projection onto the $f-H$ plane.
\label{fig:Rs_surface}}
\end{figure}

Figure~\ref{fig:Rs_surface} is a representative example of the data 
obtained, showing a surface and contour plot of $R_S(\omega,H_0)$ at 
$T=301.5$~K, a few degrees below $T_C=305.5$~K. As mentioned above, 
the field $H_0$ was applied in 
the plane of the disk-shaped sample. The sharp peak is the FMR 
absorption, which grows in intensity and moves to higher 
frequency as the field is increased. This is as we would expect from 
Eq.~(\ref{eq:wr}). At somewhat higher frequencies, the 
FMAR minimum is evident, and it also is 
seen to rise in frequency with increasing field, as predicted by 
Eq.~(\ref{eq:war}). 

\begin{figure}[htb]
\begin{center}
\leavevmode
\epsfysize=20cm
\epsfbox{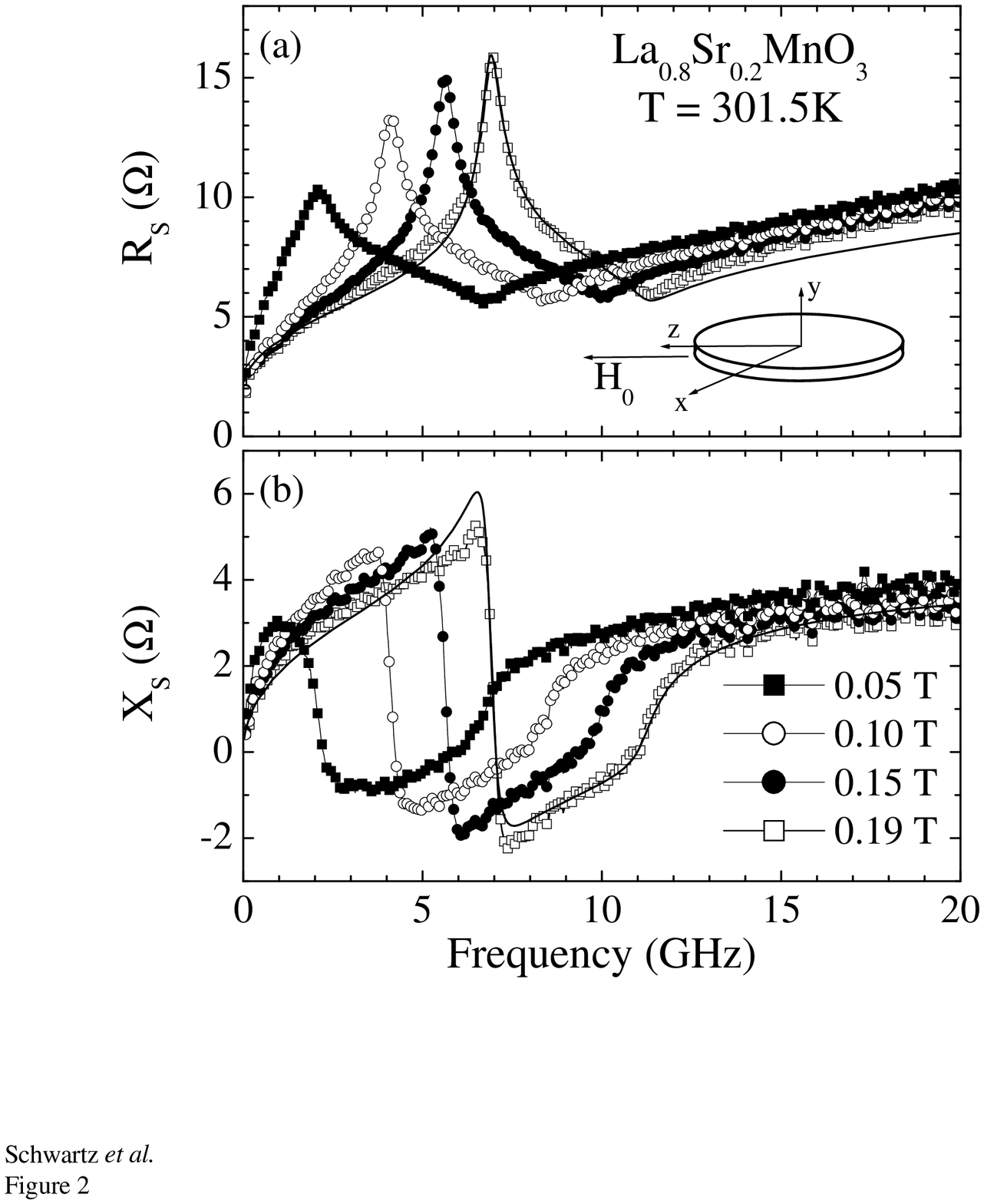}
\end{center}
\caption{The frequency dependence of the 301.5~K surface impedance 
of La$_{0.8}$Sr$_{0.2}$MnO$_3$ at finite applied field. The orientation
of the applied field $H_0$ and the axes relative to the disk-shaped
sample is shown schematically.
(a) Fixed-field cuts through the data displayed in 
Fig.~\protect\ref{fig:Rs_surface}, showing the frequency dependence of 
$R_S$ for four different values of the applied field. (b) The 
corresponding  surface reactance data at the same temperature and field 
values. The movement of both the FMR and FMAR features to higher 
frequency with increasing field is clearly seen in both sets of data. The 
thick solid line on both parts shows a fit of the model presented in 
Sec.~\protect\ref{sec:FMR} to the $\mu_0 H_0=0.19$~T data.
\label{fig:RsXs_f_H}}
\end{figure}

Fixed-field cuts through the $R_S(\omega,H_0)$ data shown in 
Fig.~\ref{fig:Rs_surface} are 
displayed in Fig.~\ref{fig:RsXs_f_H}(a); the corresponding $X_S(\omega)$ 
spectra are seen in Fig.~\ref{fig:RsXs_f_H}(b). It is clear from the latter 
that the signature of the ferromagnetic resonance is a steep drop of 
$X_S(\omega)$, with maximum slope at the resonance frequency. At the 
anti-resonance there is a less pronounced but clearly visible rise of 
$X_S$. Both of these features are predicted by the model presented in 
Sec.~\ref{sec:FMR}. From the measured resonance and anti-resonance 
frequencies, $\omega_r/2\pi=6.9$~GHz and $\omega_{ar}/2\pi=11.3$~GHz,
at an applied field of $\mu_0H_0=0.19$~T, we have uniquely determined 
the value of $M_0$ from Eqs.~(\ref{eq:wr}) and (\ref{eq:war}) as follows:
\begin{equation}
\mu_0 M_0 = \frac{1}{\gamma\omega_{ar}} (\omega_{ar}^2-\omega_r^2)
\label{eq:M0}
\end{equation}
where we have assumed that $\alpha^2\ll 1$ and therefore $\gamma^*\approx\gamma$.
In addition, as will be shown later, at this large applied field value 
$H_{loc}\ll H_0-N_z M_0$, and therefore we can also solve
Eqs.~(\ref{eq:wr}) and (\ref{eq:war}) for the static demagnetization
factor:
\begin{equation}
N_z = \frac{\mu_0 H_0 \gamma\omega_{ar} -\omega_r^2}{\omega_{ar}^2-\omega_r^2},
\label{eq:Nz}
\end{equation}
With these values fixed, $\alpha$ and $\rho_{\rm dc}$ are varied to fit the 
$Z_S(\omega,H)$ data. The thick solid lines on Fig.~\ref{fig:RsXs_f_H} are 
fits to the $\mu_0 H_0=0.19$~T data with this model and the following 
parameters: $\mu_0 M_0=0.25$~T, $\alpha=0.03$, 
$N_x^{\prime}=0$, $N_y^{\prime}=1$, $N_z=0.15$, and $\rho_{\rm dc}=60$~m$\Omega$-cm. 
The model has only been modified to account for our particular experimental geometry, 
in which $h_{\rm rf}$ is not linearly polarized, and for a small offset in 
$R_S$ and slope in $X_S$ which arise due to a contact impedance.\cite{FitCorrections} 
Similar fits can be made to the data at other temperatures and applied field 
values, and the model of Sec.~\ref{sec:FMR} clearly provides an 
excellent description of the experimental data. 
The value of the damping parameter $\alpha=0.03$ corresponds
to a linewidth $\Gamma/2\pi=\alpha\omega/2\pi$ of 0.2~GHz at the FMR frequency. 
This translates into a 70~Oe linewidth in field space, which
agrees well with the value reported earlier for this material.\cite{Lofland97b}

\begin{figure}[htb]
\begin{center}
\leavevmode
\epsfysize=13cm
\epsfbox{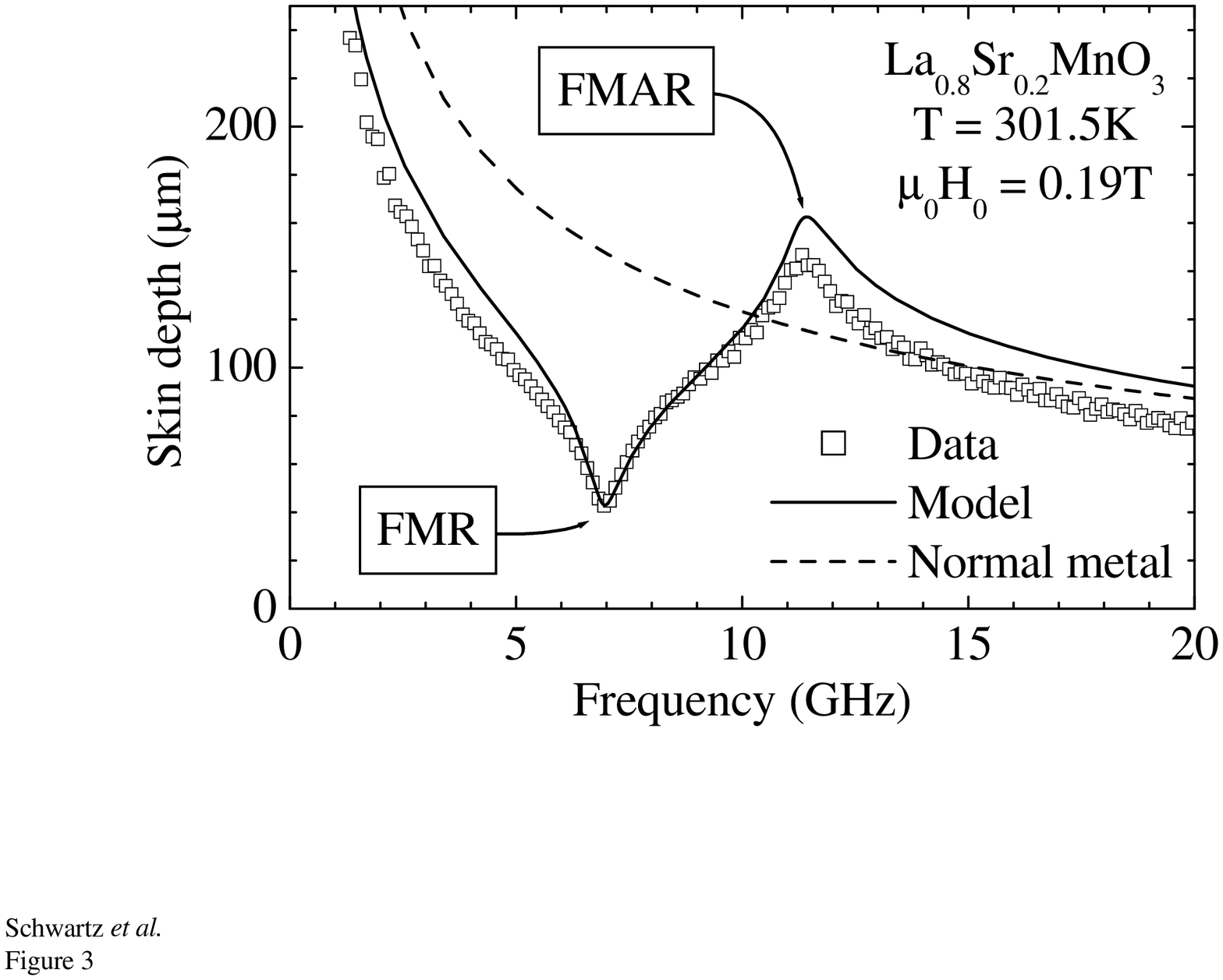}
\end{center}
\caption{Frequency dependence of the skin depth $\delta=(Re\{\hat{k}\})^{-1}$
of La$_{0.8}$Sr$_{0.2}$MnO$_3$ at 301.5~K and 0.19~T.
The open squares show the inverse of the real part of the wavevector, 
calculated from the 0.19~T data shown in Fig.~\protect\ref{fig:RsXs_f_H} 
using Eq.~(\protect\ref{eq:wavevector}). The solid line is extracted from 
the model presented in Sec.~\protect\ref{sec:FMR}, with the same parameters 
used in Fig.~\protect\ref{fig:RsXs_f_H}. The dashed line is the skin depth
for a normal, non-magnetic metal with the same resistivity value, 60~m$\Omega$-cm.
\label{fig:skin_depth}}
\end{figure}

The enhancement of $R_S$ at the FMR frequency is easily understood as an 
enhanced absorption due to a matching of the microwave frequency to the
natural precession frequency of the electronic spins within the material. The reason for
the reduction of $R_S$ at the FMAR frequency, below the value which it would 
have for a non-magnetic material with the same resistivity, is less obvious.
However, if we presume that the microwave field is a plane wave with the 
spatial and temporal dependence of the electric field given by 
$E \propto e^{i\omega t-ky}$, then it is possible to insert this 
into Maxwell's equations and solve for the wavevector $k$. 
This leads, in general, to a complex wavevector given by
\begin{equation}
\hat{k}=\sqrt{i\omega\hat{\mu}(\omega)\sigma}=\frac{\hat{Z_S}}{\rho},
\label{eq:wavevector}
\end{equation}
where we have employed Eq.~(\ref{eq:Zs}) for the second equality. As usual,
the characteristic length over which the fields decay into the material, the 
skin depth, is given by $\delta=(Re\{\hat{k}\})^{-1}$. In a non-magnetic metal 
this would just be the usual $\delta=\sqrt{2/\mu_0\omega\sigma}$, but in our case it is
modified by the frequency dependence of the permeability $\hat{\mu}$. 
Figure~\ref{fig:skin_depth} shows the frequency dependence of this skin depth
at 301.5~K and in an applied field of 0.19~T, as extracted from the data (open
squares) and the model used above to fit the surface impedance data (solid line).
The dashed line shows the skin depth for a non-magnetic metal with the same 
resistivity. What is clear in the figure is that the skin depth has a local 
maximum at the FMAR frequency, and is in fact enhanced over the non-magnetic
value. Thus at this
frequency the microwave fields penetrate farther into the sample, and the
currents are flowing in this thicker layer, thereby reducing the surface resistance.
However, it is worth noting that the thickness of this sample was approximately
800~$\mu$m, and therefore it is clear that the skin depth remains smaller than
the sample thickness even at the FMAR frequency.

\subsection{Field-dependence of $\omega_r$ and $\omega_{ar}$}
\label{sec:FrFar_H}

\begin{figure}[htb]
\begin{center}
\leavevmode
\epsfysize=20cm
\epsfbox{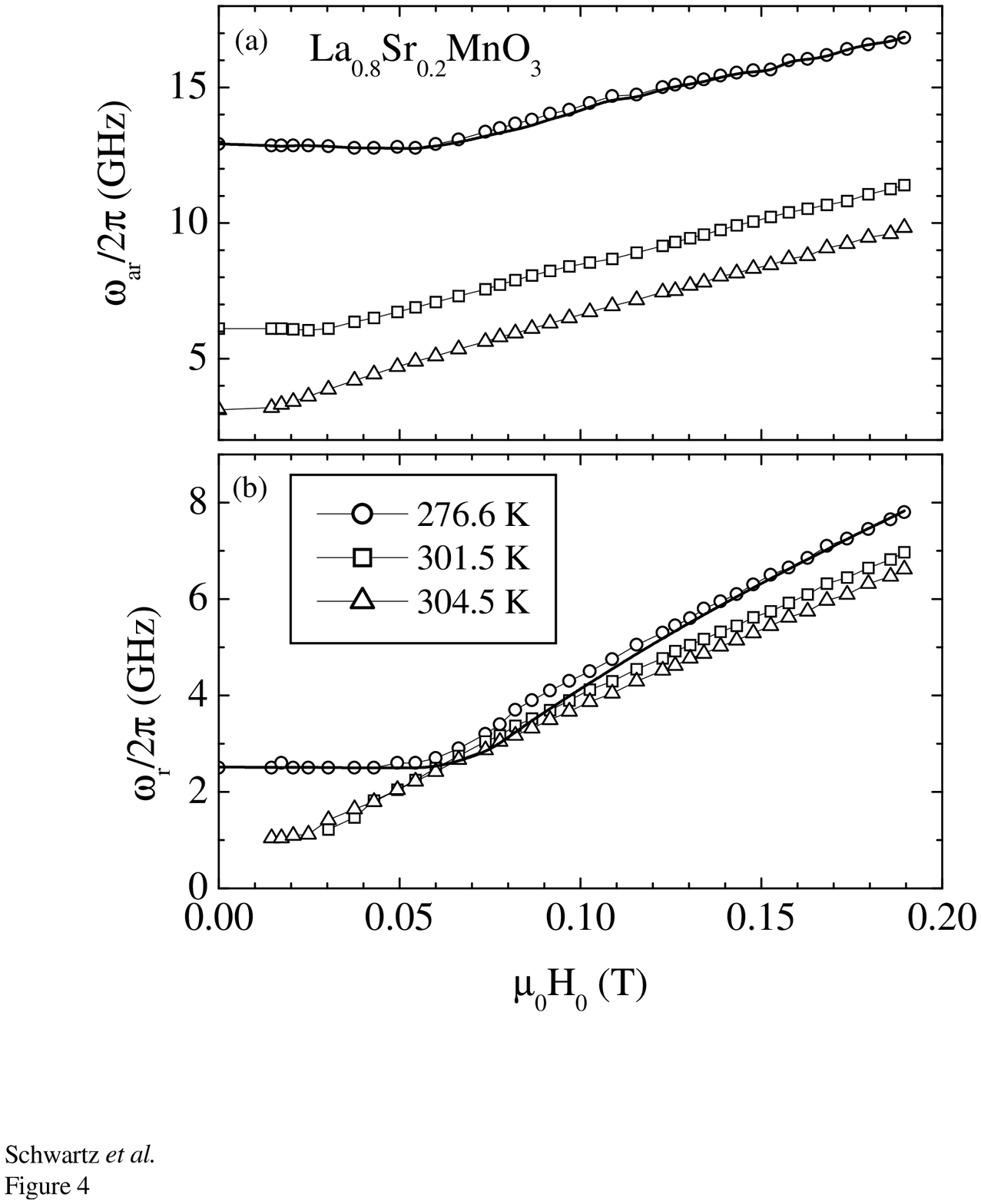}
\end{center}
\caption{Field dependence of (a) the ferromagnetic anti-resonance 
frequency and (b) the ferromagnetic resonance frequency of
La$_{0.8}$Sr$_{0.2}$MnO$_3$ at three representative
temperatures. The thick solid lines are the empirical model discussed in 
Sec.~\protect\ref{sec:FrFar_H} applied to the 276.6~K data.
\label{fig:FrFar_H}}
\end{figure}

\begin{figure}[htb]
\begin{center}
\leavevmode
\epsfysize=13cm
\epsfbox{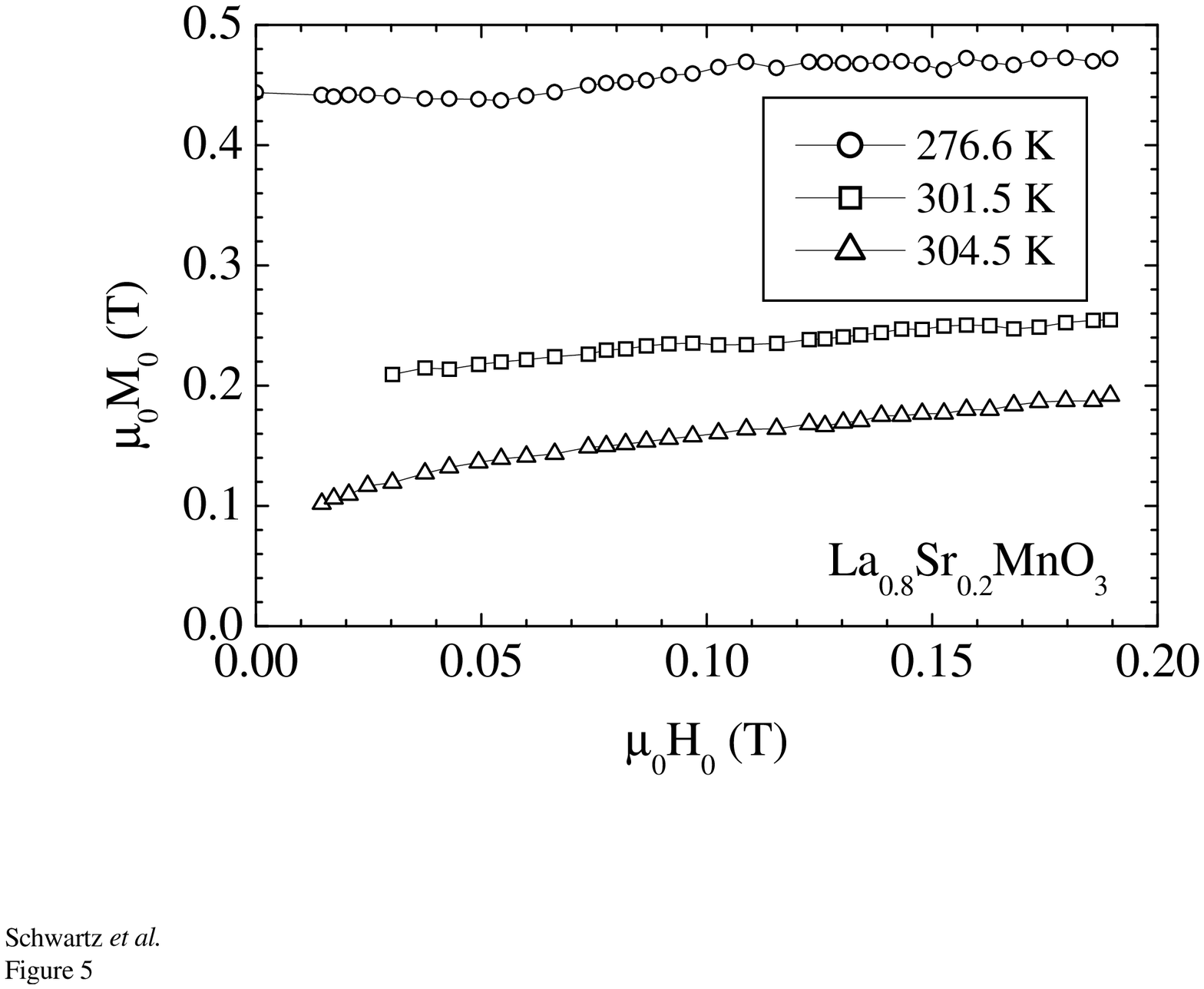}
\end{center}
\caption{The field dependence of the magnetization of 
La$_{0.8}$Sr$_{0.2}$MnO$_3$ at three representative
temperatures, as calculated from the data in Fig.~\protect\ref{fig:FrFar_H} 
using Eq.~(\protect\ref{eq:M0}).
\label{fig:M0_H}}
\end{figure}

From the data shown in Fig.~\ref{fig:Rs_surface} and similar data at other
temperatures, we can extract the field and temperature dependence of the
ferromagnetic resonance and the ferromagnetic anti-resonance frequencies. 
Figure~\ref{fig:FrFar_H} shows the magnetic field dependence of the ferromagnetic
resonance and anti-resonance frequencies at three representative temperatures, 
including 301.5~K as shown in Figs.~\ref{fig:Rs_surface} and \ref{fig:RsXs_f_H}.
As expected, both frequencies increase with increasing applied field and with
decreasing temperature. As is clear from Eq.~(\ref{eq:M0}), 
the local magnetization $M_0$
can be directly extracted from these resonance and anti-resonance frequencies.
The results of such analysis at the same three temperatures are shown in
Fig.~\ref{fig:M0_H}. As expected, once again, the magnetization is seen to grow
with decreasing temperature, and also to increase slightly with applied field.

\begin{figure}[htb]
\begin{center}
\leavevmode
\epsfysize=13cm
\epsfbox{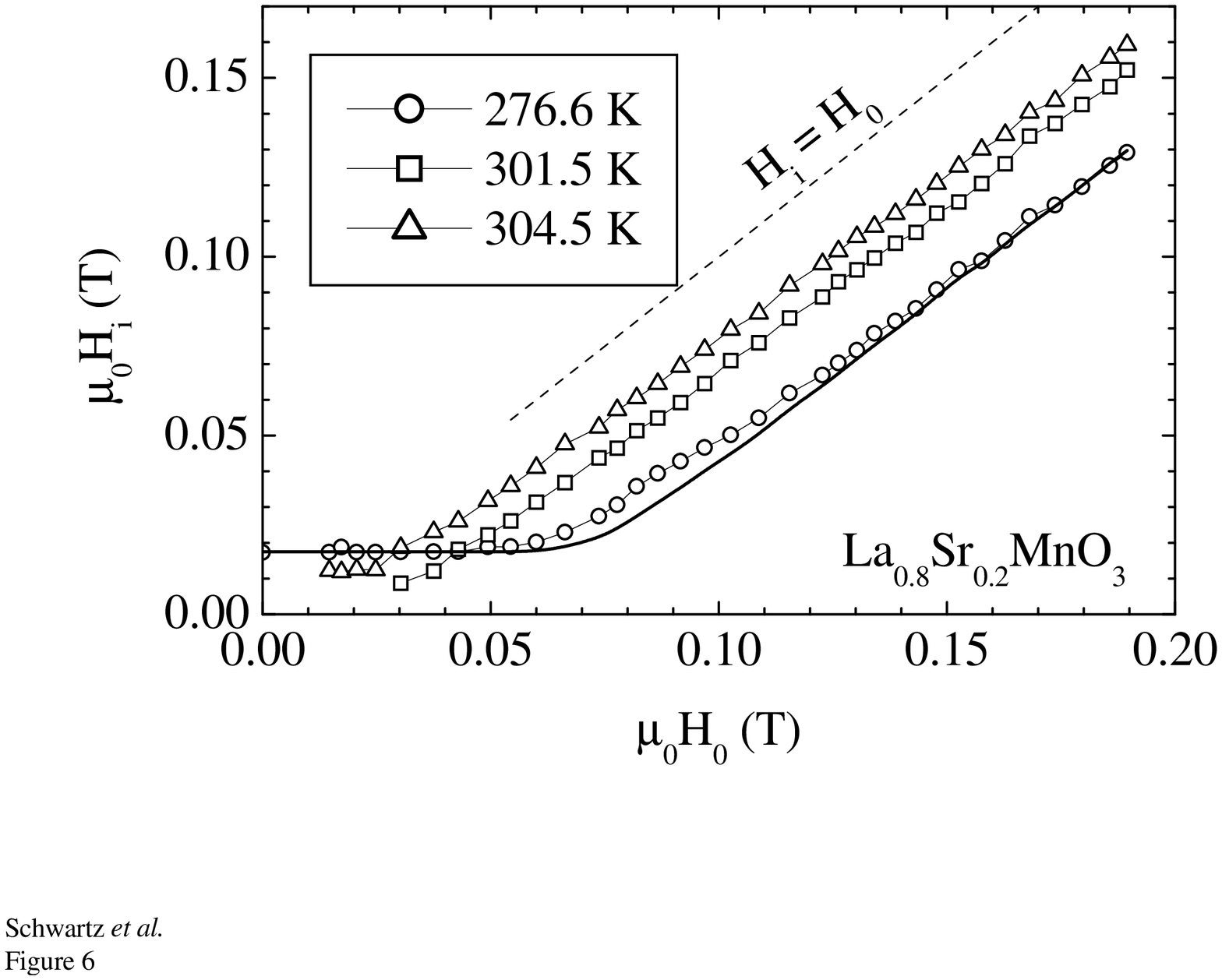}
\end{center}
\caption{The field dependence of the internal field $H_i$ at three representative
temperatures, as calculated from the data in Fig.~\protect\ref{fig:FrFar_H} 
using Eq.~(\protect\ref{eq:Hi}). The thick solid line is the empirical model 
discussed in Sec.~\protect\ref{sec:FrFar_H} applied to the 276.6~K data. 
The dashed line represents $H_i=H_0$, and it is clear that all curves
approach this slope for large values of $H_0$.
\label{fig:Hi}}
\end{figure}

As we have mentioned above, at the higher applied fields it is possible 
to use Eq.~(\ref{eq:Nz}) to determine the demagnetization factor $N_z$.
However, we have found that it is not appropriate to apply this high-field 
analysis at low fields where the FMR and FMAR frequencies become independent 
of field (see Fig.~\ref{fig:FrFar_H}). It is clear that the existence of the local
fields $H_{loc}$ cannot be neglected as the applied field approaches zero.
In fact, it is precisely these fields that allowed us to observe the FMAR, and
to extract the spontaneous magnetization, even in the absence of an applied 
field.\cite{Schwartz00}
In order to fully describe the data shown in Fig.~\ref{fig:FrFar_H}, it is
necessary to account for this field $H_{loc}$. We can solve 
Eqs.~(\ref{eq:wr}) and (\ref{eq:war}) for the total internal field
\begin{equation}
\mu_0 H_i=\frac{\omega_r^2}{\gamma\omega_{ar}}.
\label{eq:Hi}
\end{equation}
These data, again for the same three representative temperatures, are shown
as a function of the applied field in Fig.~\ref{fig:Hi}.

What we see in Fig.~\ref{fig:Hi} is that at 276.6~K for example, $H_i$ is
small and roughly constant for $H_0<0.06$~T, and then rises roughly linearly
at higher fields. This can be understood in the following way. At low applied
fields, there is reorientation of the magnetic domains within the sample
which give rise to a demagnetization field $H_d$ which exactly
cancels $H_0$. Thus in this regime $H_i=H_{loc}=0.017$~T (because of the
random orientation of $H_{loc}$ this is an average local field throughout
the sample). However, at 
$\mu_0 H_0\approx0.06$~T, the magnetic domains are all aligned, the
demagnetization field saturates at $H_d=N_z M_0$, and as the external field 
is increased it penetrates the sample. 
Therefore we presume that the demagnetization field for the data at 
276.6~K has the following form:
\begin{eqnarray}
\mu_0 H_d = \mu_0 H_0 \qquad {\rm for}\quad \mu_0 H_0 &\leq& 0.06{\rm T} \nonumber \\
\mu_0 H_d = N_z \mu_0 M_0 \quad {\rm for}\quad \mu_0 H_0 &>& 0.06{\rm T}.
\label{eq:Hdemag}
\end{eqnarray}
As we have argued earlier in deriving Eq.~(\ref{eq:Nz}), at large field
values $H_{loc}$ can be ignored. This is due in part to the fact that
$H_{loc}\ll H_0$, but also to the fact that $H_{loc}$ is randomly oriented within
the sample. Therefore for fields $H_0\gg H_{loc}$ we would expect that $H_{loc}$ 
would simply broaden the FMR and FMAR lines and not shift them. If we calculate
the total internal field $H_i$ using the known applied field, the 
demagnetization field given by Eq.~(\ref{eq:Hdemag}), and an $H_{loc}$ which
has a constant magnitude of 0.17~T but is averaged over all orientations, then
we arrive at the solid line on Fig.~\ref{fig:Hi}. It is clear that this 
provides an excellent description of the dependence of the total internal
field on the applied external field.

Finally, using this empirical form for $H_i$, and the magnetization data
shown in Fig.~\ref{fig:M0_H}, it is possible to model the complete field
dependence of the ferromagnetic resonance and anti-resonance frequencies 
with Eqs.~(\ref{eq:wr}) and (\ref{eq:war}). This is shown, again only
at 276.6~K, by the solid lines on Fig.~\ref{fig:FrFar_H}, and it is clear that it
provides an excellent description of the data throughout the entire 
range of applied fields. The same analysis can be carried out at other
temperatures in order to obtain a full description of all of the data.

\section{Conclusions}

By measuring the broadband frequency dependence of the surface impedance
of La$_{0.8}$Sr$_{0.2}$MnO$_3$ we have shown that the Landau-Lifshitz-Gilbert
expression for the dynamic susceptibility provides an excellent description
of the complete microwave spectra, including the ferromagnetic resonance and
anti-resonance features. In addition, we have performed a quantitative 
analysis of the field dependence of these features, and
have shown that they can be modeled extremely well with rather simple
expressions for the field dependence of the magnetization, the demagnetization
field, and the local internal fields. To our knowledge, this is the most
extensive study of the frequency, field, and temperature dependence of the
microwave properties of a ferromagnet.

In addition, we would like to stress that in contrast to conventional
magnetization measurements, the technique presented here
is sensitive to the average {\em local} magnetization of the material. 
At each point in the sample, the
local magnetization is interacting with the local static and dynamic 
magnetic fields to produce the dramatic FMR and FMAR features,
which are then averaged over the entire sample. As we have reported
earlier, this fact allowed us to measure the spontaneous magnetization
of this material in the absence of any applied magnetic field.\cite{Schwartz00}
Conventional magnetization measurements require the application of an
applied field in order to produce a net magnetization in the sample.
Finally, the broadband nature of the surface impedance measurements
presented here have allowed us to determine this local magnetization
of La$_{0.8}$Sr$_{0.2}$MnO$_3$ over wide magnetic field and temperature
ranges.

\acknowledgements
We would like to thank Y. Mukovskii and colleagues at the Moscow State Steel 
and Alloys Institute for growing the samples used in this study. We also thank
S. Bhagat and S. Lofland for providing the samples to us, and for many
useful and interesting discussions about these materials. This work was supported 
by NSF Grant No. DMR-9624021, the University of Maryland NSF-MRSEC Grant No. 
DMR-0080008, and the Maryland Center for Superconductivity Research.

\end{multicols}
\widetext

\newpage

\newpage

\newpage

\newpage

\newpage

\newpage


\begin{thebibliography}{99}
\bibitem{CuNote} The low resistivity of Cu, and it's small temperature 
dependence over the temperatures of interest in this study, make this
an extremely good assumption.
\bibitem{FitCorrections} In the coaxial geometry, the microwave 
magnetic field lines are circular, and therefore $h_{\rm rf}$ is not 
everywhere orthogonal to $H_0$, as was assumed by the model presented
in Sec.~\protect\ref{sec:FMR}. In order to account for this we have 
calculated the effective impedance of the sample by assuming that the
susceptibility varies as $\cos^2\theta$ and performing the angular 
integration. In addition, we find that there is a small impedance 
mismatch between the coaxial cable and the sample which we have modeled 
as a parallel resistance and capacitance, and included in the effective 
impedance. It is this effective impedance which is fit to the data in Fig.~\protect\ref{fig:RsXs_f_H}.
\end{thebibliography}
\end{document}